\documentclass[preprint,5p,twocolumn]{elsarticle}

\usepackage[utf8]{inputenc}
\usepackage{dcolumn}
\usepackage{amsmath}
\usepackage{amsfonts,amssymb,bm,tensor,braket}
\usepackage{xcolor}
\usepackage{autobreak}
\usepackage{graphicx}
\usepackage{threeparttable,array,booktabs}
\usepackage[referable]{threeparttablex}
\usepackage{makecell, multirow}
\usepackage{tabularx}
\usepackage{setspace}
\usepackage[varg]{txfonts}
\usepackage{caption}
\usepackage[colorlinks=true,allcolors=cyan]{hyperref}
\usepackage[nameinlink,noabbrev]{cleveref}

\bibpunct{\color{cyan}[}{\color{cyan}]}{,}{n}{}{;}

\journal{PLB, published as Phys. Lett. B 819 (2021) 136443, \url{https://doi.org/10.1016/j.physletb.2021.136443}}

\begin{document}
\hypersetup{allcolors=cyan}

\begin{frontmatter}



\newcommand{\e}{{\mathrm{e}}}
\renewcommand{\i}{{\mathrm{i}}}
\renewcommand{\deg}{^\circ}

\newcommand*{\PKU}{School of Physics and State Key Laboratory of Nuclear Physics and Technology, Peking University, Beijing 100871, China}
\newcommand*{\CHEP}{Center for High Energy Physics, Peking University, Beijing 100871, China}
\newcommand*{\CIC}{Collaborative Innovation Center of Quantum Matter, Beijing, China}

\title{Light Speed Variation in a String Theory Model for Space-Time Foam}

\author[a]{Chengyi Li}
\author[a,b,c]{Bo-Qiang Ma\corref{cor1}}

\address[a]{\PKU}
\address[b]{\CHEP}
\address[c]{\CIC}
\cortext[cor1]{Corresponding author \ead{mabq@pku.edu.cn}} 

\begin{abstract}
We revisit a supersymmetric string model for space-time foam, in which bosonic open-string states, such as photons, can possess quantum-gravity-induced velocity fluctuations in vacuum. We argue that the suggestion of light speed variation with lower bound from gamma-ray burst photon time delays can serve as a support for this string-inspired framework, through connecting the experimental finding with model predictions. We also derive the value of the effective quantum-gravity mass in this framework, and give a qualitative study on the model-dependent coefficients. Constraints from birefringent effects and/or photon decays, including the novel $\gamma$-decay constraint obtained here from the latest Tibet AS$\gamma$ near-PeV photon, are also found to be consistent with predictions in such a quantum-gravity scheme. Future observation that can testify further the theory is suggested.
\end{abstract}

\begin{keyword}
light speed variation\sep gamma-ray burst \sep stringy space-time foam\sep Lorentz invariance violation



\end{keyword}

\end{frontmatter}

Pursuing a unified description of all the fundamental forces has motivated various attempts towards quantum gravity (QG). The most promising framework for a compelling approach to gravity quantization up to now is string theory. In certain string models~\cite{Ellis:1999uh,Ellis:2000sx}, quantum-gravitational fluctuations in the space-time background - `space-time foam' - can be treated as point-like D0-branes in the bulk space-time~\cite{Ellis:2005qa}, resulting in the speculation that the vacuum might behave essentially like a dispersive medium that could violate Lorentz invariance. Such a symmetry breakdown may lead to observable consequences for the light propagation. Especially, the light velocity \textit{in vacuo}, postulated to be a constant $c$ in Einstein's relativity, might suffer tiny modifications due to Lorentz violation~(LV) effects at the Planck scale $E_{\mathrm{Pl}}\equiv\sqrt{\hslash c^{5}/G_{N}}\simeq 10^{19}~\mathrm{GeV}$. It has been proposed~\cite{AmelinoCamelia:1996pj,AmelinoCamelia:1997gz} that distant astrophysical objects of high-energy photon emissions, \textit{e.g.}, $\gamma$-ray bursters (GRBs), can provide crucial tests of such LV effects. Ellis \textit{et al}.~\cite{Ellis:1999sd} first analyzed the GRB data aiming to detect QG-motivated deformation of light velocity. Lately a series of studies~\cite{Shao:2009bv,Zhang:2014wpb,Xu:2016zxi,Xu:2016zsa,Xu:2018ien,Liu:2018qrg,Amelino-Camelia:2016ohi} on \textit{Fermi}-LAT GRB photons suggest a subluminal light speed variation of the form $1-v/c=\mathcal{E}/E_{\mathrm{LV}}$ at the Lorentz-violation scale of $3.6\times 10^{17}~\mathrm{GeV}$~\cite{Xu:2016zxi,Xu:2016zsa,Xu:2018ien}. In this paper, we point out that this previously suggested constraint on light speed variation can serve as a support for the D-brane/string-inspired model for space-time foam, through connecting the experimental finding with such a theory, in agreement with other astrophysical constraints currently available. We also suggest to testify this theory further, using future observations.

The intuitive picture of \textit{space-time foam} goes all the way back to Wheeler~\cite{Wheeler:1998gb} who first noted that there would be highly curved quantum fluctuations in the space-time continuum with $\Delta E\sim E_{\mathrm{Pl}}$ at short time and small spatial scales $\Delta t\sim\Delta x\sim \hslash/E_{\mathrm{Pl}}$, at which QG effects become relevant. Within a Liouville-inspired stringy model for space-time foam~\cite{Ellis:2004ay,Ellis:2008gg,Li:2009tt}, our Universe is described as a D(irichlet)3-brane, roaming in the bulk space-time of a high-dimension cosmology, wherein quantum-gravitational fluctuations are represented as D0-brane~(D-particle) defects, which are allowed in Type-IA superstrings~\cite{Schwarz:1999xj}. These D-particle defects appear to an observer on the brane world as space-time foam~(termed as `D-foam') that `flashes' on and off. Standard-model particles are viewed as~(open-)string excitations with their ends attached on the brane. Their interaction with D-defects then modifies the canonical dispersion relations of certain particles to include QG corrections that depend on their energies.

Here we work in an anisotropic D-foam model~\cite{Ellis:2004ay}, in which collisions of photon open-strings with D-particles cause the latter to recoil, inducing a Finsler-like target-space metric
\begin{equation}\label{eq:1}
G_{\alpha\beta}=\eta_{\alpha\beta}+h_{\alpha\beta},\quad h_{0i}=U_{i}\ll 1,
\end{equation}
where $U_{i}=(g_{s}/M_{s})\Delta p_{i}=\Delta p_{i}/M_{D}$ is the velocity of the recoiling D-particle following scattering by the photon through a momentum transfer $\Delta p_{i}\equiv\lambda p_{i}~(\lambda<1)$. Here $g_{s}<1$ is the string coupling, $M_{s}$ is the string scale, and the D-particle mass $\sim M_{s}/g_{s}\equiv M_{D}$. Such a metric fluctuation~\labelcref{eq:1} then affects the dispersion relation of photons through the on-shell relation\footnote{Natural units where $\hslash, c\equiv 1$ are adopted from now on.}
\begin{equation}\label{eq:2}
p^{\alpha}p^{\beta}G_{\alpha\beta}=-\mathcal{E}^{2}+\mathbf{p}^{2}+2g_{s}\frac{\lambda}{M_{s}}\mathcal{E}\mathbf{p}^{2}=0,
\end{equation}
where $(\mathcal{E},\mathbf{p})$ indicates the four-momentum of the photon. Due to stochastic captures of the photon by multiple D-particle defects, there would be an \textit{average} background recoil-velocity field $\langle\!\langle U_{i}\rangle\!\rangle_{D}\neq 0$ over the collection of D-particles encountered by the photon\footnote{To make it clear that we denote this type of averaged vacuum expectation values with the notation: $\langle\!\langle\cdot\!\cdot\!\cdot\rangle\!\rangle_{D}$, since the averaging, which depends in general on the details of the foam, cannot be performed quantitatively.}. The dispersion relation~\labelcref{eq:2} then leads~(on average) to a deformed velocity of light in the D-brane foam\footnote{For simplicity, we omit the averaging notation for $p_{i}$ and $\mathcal{E}$, so keep in mind that these quantities as concerned in the model denote such averages.}
\begin{equation}\label{eq:3}
c_{g}=1-2g_{s}\frac{\zeta_{D}\lvert\mathbf{p}\rvert}{M_{s}}\simeq1-\mathcal{O}\left(g_{s}\frac{n_{D}\mathcal{E}}{M_{s}}\right),\quad \langle\!\langle\lambda\rangle\!\rangle_{D}=\zeta_{D}>0,
\end{equation}
where, to leading order, $\lvert\mathbf{p}\rvert\simeq\mathcal{E}$, and the coefficient $\zeta_{D}$ depends on the density of the D-particle $n_{D}$, which is essentially a free parameter, leaving the fact that the (stringy) QG scale $M_{\mathrm{QG}}=M_{D}/2\zeta_{D}$ is free to vary and, thus, needs to be determined by experiment\footnote{We assume here a \textit{uniform} D-particle background, \textit{i.e.}, $\zeta_{D}$ is redshift-$z$-independent. Otherwise it could vary with the cosmological epoch~\cite{Mavromatos:2009mj}, then the QG scale could depend also on $z$: $M_{\mathrm{QG}}\mapsto\widetilde{M_{\mathrm{QG}}}(z)\sim n_{D}(z)/M_{s}$.}. The averaged dispersion relation leads also to a non-trivial refractive index, $\eta-1$, of the form
\begin{equation}\label{eq:4}
\eta=1+\mathcal{O}\left(\zeta_{D}\mathcal{E}/M_{D}\right)>1.
\end{equation}

Some important remark is in order here, to make it easier as we compare the experiment with theoretical predictions in the following:~\textbf{(i)} As seen from Eq.~\labelcref{eq:3}, the above model motivates a light velocity fluctuation that grows \textit{linearly} with the energy of that light, and~\textbf{(ii)} such a modified velocity corresponds also to a refractive index~\labelcref{eq:4} which is \textit{subluminal}. The sign of the velocity variation $\Delta c_{g}\equiv c_{g}-1$ was determined by the fact that the underlying string theory naturally prevents superluminal propagation. Such result can be understood physically also from the fact that the metric perturbation~\labelcref{eq:1} is $\langle\!\langle\mathbf{U}\rangle\!\rangle_{D}$-dependent, and is consequently energy-dependent. This affects photons on their way of propagation in such a way that more energetic photons would `see' more foam, and thus are decelerated with respect to low-energy ones, and~\textbf{(iii)} as a result, photons are \textit{stable}.~\textbf{(iv)} More importantly, the light velocity~\labelcref{eq:3} is seen to be \textit{helicity-symmetric}, therefore the vacuum is \textit{not} birefringent.

It was first proposed by Amelino-Camelia \textit{et al}.~\cite{AmelinoCamelia:1996pj,AmelinoCamelia:1997gz} that distant celestial sources of high energy photons, such as GRBs, pulsars and active galactic nuclei (AGNs), can play a crucial role in probing such QG-motivated deformation of light velocity. Ellis \textit{et al}.~\cite{Ellis:1999sd} presented the first analysis on GRB data of photons to detect QG light speed variance. Here we indicate in the following that a recently suggested Lorentz-violating picture~\cite{Zhang:2014wpb,Xu:2016zxi,Xu:2016zsa,Xu:2018ien,Liu:2018qrg,Amelino-Camelia:2016ohi} emerging from \textit{Fermi}-LAT GRB time-delay data can serve as a support to the stringy space-time foam model discussed above.

In Refs.~\cite{Xu:2016zxi,Xu:2016zsa,Xu:2018ien}, utilizing a phenomenological Lorentz-violation modified dispersion relation
\begin{equation}\label{eq:5}
\mathcal{E}^{2}=\mathbf{p}^{2}\left[1-s_{n}\left(\frac{\lvert\mathbf{p}\rvert}{E_{\mathrm{LV}, n}}\right)^{n}\right],
\end{equation}
the authors get the propagation velocity for photons as
\begin{equation}\label{eq:6}
v=1-s_{n}\frac{n+1}{2}\left(\frac{\mathcal{E}}{E_{\mathrm{LV},n}}\right)^n,
\end{equation}
where $n=1$ or $2$ corresponds to linear or quadratic energy dependence, $s_{n}=\pm1$ is a sign factor of LV correction, and $E_{\mathrm{LV},n}$ is the $n$th-order Lorentz-violation scale. Analyses on GRB photons performed by the authors lead to a suggestion of a \textit{subluminal}~(\textit{i.e.}, $s=+1$) light speed variation\footnote{Compared to the quadratic~($n=2$) case, as illustrated in Ref.~\cite{Xu:2016zxi}, the linear~($n=1$) LV case is more favored by the data. In the case of $n=1$ the redundant subscripts $n$ in Eqs.~\labelcref{eq:5} and~\labelcref{eq:6} are hereafter omitted.} in vacuum of the form $v(\mathcal{E})=1-\mathcal{E}/E_{\mathrm{LV}}^{(\mathrm{sub})}$ with $E_{\mathrm{LV}}^{(\mathrm{sub})}=(3.60\pm 0.26)\times 10^{17}~\mathrm{GeV}$, the first-order Lorentz-violation scale, which is close to the Planck energy scale $E_{\mathrm{Pl}}\simeq1.22\times10^{19}~\mathrm{GeV}$. Due to the fact that there could be uncertainties coming from unknown source-intrinsic mechanisms or cosmological matter effects when analyzing the data, we should interpret this previous experimental finding as a lower bound on \textit{subluminal} Lorentz violation with
\begin{equation}\label{eq:7}
E_{\mathrm{LV}}^{(\mathrm{sub})}\gtrsim 3.60\times 10^{17}~\mathrm{GeV}.
\end{equation}
It is worth noting that such a Lorentz violation bound is consistent with various constraints from high-energy $\gamma$-ray observations of pulsars~\cite{Zitzer:2013gka}, AGNs~\cite{Romoli:2017zuc,Abdalla:2019krx,Li:2020uef} as well as GRBs~\cite{Ellis:1999sd,Ellis:2005wr,Bolmont:2006kk,Chang:2015qpa,Bernardini:2017tzu,Pan:2020zbl}, and it is also compatible with the strongest robust limit to date from a recent study~\cite{Ellis:2018lca} on 8 \textit{Fermi}-LAT GRBs. We need to mention that there can be severer bounds to $E_{\mathrm{LV}}$ from other time-delay studies~(see, \textit{e.g.}, Ref.~\cite{Acciari:2020kpi}), thus more observations are needed to further verify the proposed light speed variation in previous works.

Notice that, in the above phenomenological LV picture, one of the most interesting features revealed from data analyses is that multi-GeV events from various different samples of GRBs line up surprisingly to form a `mainline', see Refs.~\cite{Xu:2016zxi,Xu:2016zsa,Xu:2018ien} and~\cite{Amelino-Camelia:2016ohi}, leading towards the strong indication of a \textit{linear energy dependence} in the speed of light, $\lvert v-1\rvert\sim\mathcal{O}\left(\mathcal{E}/E_{\mathrm{LV}}^{(\mathrm{sub})}\right)$, rather than a quadratic~(or higher-order) dependence of the photon energy, with $E_{\mathrm{LV}}^{(\mathrm{sub})}$ characterizing such a linear suppression of LV exceeding a few $\times$ $10^{17}~\mathrm{GeV}$, \textit{i.e.}, Eq.~\labelcref{eq:7}. On the other hand, analyses on data lead also to a positive sign factor $s=+1$ for this linear LV
correction, implying that high-energy photons propagate more slowly than their low-energy counterparts, corresponding to a \textit{subluminal photon propagation} in cosmic space. If we consider these results in terms of the photon phase velocity $\sim\mathcal{E}/\lvert\mathbf{p}\rvert=1/\eta$, taking into account Eq.~\labelcref{eq:5}, with
\begin{equation}\label{eq:8}
\eta=1+\frac{1}{2}s_{n}\left(\frac{\mathcal{E}}{E_{\mathrm{LV},n}}\right)^{n},
\end{equation}
the light speed variation with $E_{\mathrm{LV}}^{(\mathrm{sub})}\gtrsim 3.6\times 10^{17}~\mathrm{GeV}$ from GRBs just indicates that the vacuum exhibits non-trivial `optical' properties regarding an index of refraction for the photon
\begin{equation}\label{eq:8a}
\eta\simeq 1+\beta\left(\frac{\mathcal{E}}{\mathrm{GeV}}\right)>1,\ \ \ \beta\lesssim\mathcal{O}\left(10^{-18}\right),
\end{equation}
where $\beta$ is a positive numerical factor determined by Eq.~\labelcref{eq:7}. Such phenomenon~\labelcref{eq:8a} is exactly what the stringy space-time foam model has predicted, see Eq.~\labelcref{eq:4}. In view of the above discussion, we emphasize again that the current photon time-delay data do favor the suggestion that photons should possess a modified \textit{subluminal} propagation speed with Lorentz-violating terms that vary \textit{linearly} with their energies, and now it becomes clear that these two distinct LV properties that photon must satisfy, as observed from GRB time delays~\cite{Shao:2009bv,Zhang:2014wpb,Xu:2016zxi,Xu:2016zsa,Xu:2018ien,Liu:2018qrg}, indeed agree with the predictions~\textbf{(i)} and~\textbf{(ii)}~(and consequent~\textbf{(iii)}) of the D-particle string foam model just revisited.

With the established correspondence of the velocity from the generalized Lorentz-violation modified dispersion relation in Eq.~\labelcref{eq:6} to the velocity from the D-particle foam model in Eq.~\labelcref{eq:3}, we can relate $M_{\mathrm{QG}}$ to the Lorentz-violation scale $E_{\mathrm{LV}}^{(\mathrm{sub})}$ from GRBs. Then we arrive at a relation
\begin{equation}\label{eq:9}
M_{\mathrm{QG}}\simeq E_{\mathrm{LV}}^{(\mathrm{sub})}\gtrsim 3.60\times 10^{17}~\mathrm{GeV},
\end{equation}
or, equivalently we have
\begin{equation}\label{eq:10}
\frac{g_{s}\zeta_{D}}{M_{s}}\simeq\frac{1}{2E_{\mathrm{LV}}^{(\mathrm{sub})}}\lesssim 1.39\times 10^{-18}~\mathrm{GeV}^{-1}.
\end{equation}
The result~\labelcref{eq:9} leads to a lower bound on the effective quantum-gravity scale $\gtrsim 10^{17}~\mathrm{GeV}$ for the string/D-particle foam model. This is very close to the expectation that this scale is taken to be very large, typically an order of magnitude or so below the Planck scale, $M_{\mathrm{QG}}\lesssim M_{\mathrm{Pl}}=\mathcal{O}\left(10^{18}-10^{19}\right)~\mathrm{GeV}$, where $M_{\mathrm{Pl}}$ may be referred to as the~(reduced) Planck mass $\sim 2\times 10^{18}~\mathrm{GeV}$. It is also worth mentioning that our result~\labelcref{eq:9} is very similar to that obtained recently in Ref.~\cite{Ellis:2018lca}, where the first-order LV mass scale was studied using statistical estimators, with lower bound exceeding either $8.4\times 10^{17}~\mathrm{GeV}$ or $2.4\times 10^{17}~\mathrm{GeV}$. Given that $\zeta_{D}\lesssim\mathcal{O}\left(1\right)$, we can further deduce the constraint
\begin{equation}\label{eq:11}
\frac{M_{s}}{g_{s}}<7.20\times 10^{17}~\mathrm{GeV},
\end{equation}
which is compatible with the natural assumption~\cite{Mavromatos:2009mj} for the value $M_{D}=M_{s}/g_{s}\sim 10^{19}~\mathrm{GeV}$ of the mass of the foam defect. Meanwhile, we find that our result leads to a rough estimate about the scale $M_{s}\sim 3.60\times 10^{17}~\mathrm{GeV}$, which gives a string mass nearly $0.15\times M_{\mathrm{Pl}}$ if the quantity $2\zeta_{D}g_{s}$ is of order $1$. While in general, taking Eq.~\labelcref{eq:10} into account, we can get
\begin{equation}\label{eq:12}
M_{s}\gtrsim 7.20\times 10^{17}\zeta_{D}g_{s}~\mathrm{GeV}.
\end{equation}
If $\zeta_{D}g_{s}\sim\mathcal{O}\left(1\right)$ and then an intriguing estimation of $M_{s}$ is at the scale of a few $\times$ $10^{17}~\mathrm{GeV}$, which is consistent with the expectation in modern string theory that $M_{s}\neq M_{\mathrm{Pl}}$ in general. Whilst usually it is assumed in the string foam model that the combination $\zeta_{D}g_{s}$ is (much) lower than unity, then the string scale $M_{s}$ is roughly not higher than $10^{16}~\mathrm{GeV}$ and it can be as low as a few $\mathrm{TeV}$, which could be attained if $\zeta_{D}\ll 1$ or $g_{s}\ll 1$. In this case, the result may be compatible with several schemes of low-scale strings~\cite{Antoniadis:1999rm,Antoniadis:2000vd} in which the mass scale $M_{s}$ can be decreased arbitrarily low, though it cannot be lower than a few tens of TeV since if this were the case we should have already observed fundamental strings at the Large Hadron Collider. In addition, we can also obtain analogous estimations of the string length $\ell_{s}=1/M_{s}$ and the universal Regge slope $\alpha^{\prime}$ as
\begin{align}\label{eq:13and14}
\ell_{s}\lesssim&\ 1.39\times 10^{-18}\frac{1}{\zeta_{D}g_{s}}~\mathrm{GeV}^{-1},\\
\alpha^{\prime}\lesssim&\ 1.93\times 10^{-36}\frac{1}{\zeta_{D}^{2}g_{s}^{2}}~\mathrm{GeV}^{-2}.
\end{align}
For fundamental string length $\ell_{s}$, our estimate~\labelcref{eq:13and14} further indicates an upper bound $\sim 1.69\times 10^{-19}\frac{1}{\zeta_{D}g_{s}}~\mathrm{GeV}^{-1}\left(\frac{\ell_{\mathrm{Pl}}}{10^{-28}~\mathrm{eV}^{-1}}\right)\approx 1.7\frac{\ell_{\mathrm{Pl}}}{\zeta_{D}g_{s}}\sim\ell_{\mathrm{Pl}}/n_{D}g_{s}$, with the Planck length $\ell_{\mathrm{Pl}}\simeq 10^{-35}~\mathrm{m}$.

From the above discussion, we conclude eventually that the finding of the light speed variation of the form $v=1-\mathcal{E}/E_{\mathrm{LV}}^{(\mathrm{sub})}$ from GRB photons is compatible with the string-foam motivated deformation of light velocity~\labelcref{eq:3} with $M_{\mathrm{QG}}\gtrsim 10^{17}~\mathrm{GeV}$, approaching the Planck mass, through relating the stringy QG mass scale to the linear-order Lorentz-violation scale $E_{\mathrm{LV}}^{(\mathrm{sub})}$ from GRBs.
This fact thus can serve as a support to this type of models of string theory. Some severer limits to $E_{\mathrm{LV}}$ from other photon time-delay studies, as mentioned above, might be also consistent with this stringy QG theory, as long as linear suppressions of subluminal LV photon propagation is suggested from pertinent analyses on data.

However there are still several classes of complementary~(astrophysical) results, which need to be assessed in our attempt to account for the experimental finding in Refs.~\cite{Xu:2016zxi,Xu:2016zsa,Xu:2018ien} with the theory discussed here. These results are those strong constraints that come from Lorentz-violating photon decays as well as birefringence effects. Whereas, we shall point out below that these much stringent limits on $E_{\mathrm{LV}}$ are also consistent with the D-particle space-time foam framework.

To see this clearly, we focus first on strong constraints based on photon decay, $\gamma\rightarrow e^{+}+e^{-}$, which is normally forbidden but may be kinematically allowed, due to superluminal Lorentz violation~(\textit{i.e.}, $s=-1$) in Eq.~\labelcref{eq:5}, to take place once the photon energy is above the decay threshold
\begin{equation}\label{eq:15}
\frac{4m_{e}^{2}}{\mathcal{E}\left(\mathcal{E}^{2}-4m_{e}^{2}\right)}\leqslant 1/E_{\mathrm{LV}}^{(\mathrm{sup})},
\end{equation}
which reads $\mathcal{E}\gtrsim\left(4m_{e}^{2}E_{\mathrm{LV}}^{(\mathrm{sup})}\right)^{1/3}\equiv\mathcal{E}_{\mathrm{th}}$, after taking the electron rest mass $m_{e}\simeq511~\mathrm{keV}\ll \mathcal{E}$ into account\footnote{We have neglected any Lorentz violation effect for electrons, since it has been severely constrained by the absence of the vacuum Cherenkov effect.}. Since photon decay in flight from the source would lead to a hard cutoff in the astrophysical spectra of gamma-rays, a lower bound for superluminal LV energy scale $E_{\mathrm{LV}}^{(\mathrm{sup})}$ is then provided directly by any observed cosmic photon with energy greater than the threshold $\mathcal{E}_{\mathrm{th}}$. Specifically, from Eq.~\labelcref{eq:15} we have
\begin{equation}\label{eq:16}
E_{\mathrm{LV}}^{(\mathrm{sup})}\gtrsim9.57\times 10^{23}~\mathrm{eV}~\left(\frac{\mathcal{E}}{\mathrm{TeV}}\right)^{3}.
\end{equation}
Actually, the observations of multi-TeV photons from the Crab Nebula have already cast stringent limits~\cite{Martinez-Huerta:2016azo} with the relevant scale constrained to the level $E_{\mathrm{LV}}^{(\mathrm{sup})}\gtrsim 10^{20}~\mathrm{GeV}$. A tighter result comes from recent observations of $\gamma$-rays around $100~\mathrm{TeV}$ with HAWC~\cite{Albert:2019nnn}. Therein a $2\sigma$ constraint is deduced to be $E_{\mathrm{LV}}^{(\mathrm{sup})}>2.22\times 10^{22}~\mathrm{GeV}$. However, thanks to the most recent detection of Galactic $\gamma$-rays with energies up to nearly 1~PeV reported by the Tibet AS$\gamma$ collaboration~\cite{Amenomori:2021gmk}, here we can obtain the strongest constraint to date, by inserting the highest energy of the observed photon, $\mathcal{E}_{\max}=957(^{+166}_{-141})~\mathrm{TeV}$ into Eq.~\labelcref{eq:16},
\begin{equation}\label{eq:17}
E_{\mathrm{LV}}^{(\mathrm{sup})}\gtrsim 8.39^{+5.17}_{-3.19}\times 10^{23}~\mathrm{GeV},
\end{equation}
over about 68000 times the Planck energy, hence we have an improvement of 1 to 3 orders of magnitude over previous limits to superluminal photon decay. Obviously, these strong constraints just indicate that there is \textit{no} such a cutoff compatible with Lorentz-violating $\gamma$-decay in the measured spectra, such that those very-high-energy $\gamma$-ray photons of TeV energies could travel to the Earth without undergoing a rapid self-decay on their way of propagation. This is exactly what the D-particle foam model has expected, that is the prediction~\textbf{(iii)}, as mentioned. Indeed, in the D-brane space-time foam photon open-strings \textit{only} possess \textit{subluminal} dispersion relations, as seen from Eqs.~\labelcref{eq:3} and~\labelcref{eq:4} and/or characterized by $\zeta_{D}>0$, without any suggestion of superluminal propagation. Photons are therefore \textit{stable} (\textit{i.e.}, do \textit{not} decay), and hence using the results based on superluminal dominant phenomena~(\textit{e.g.}, $\gamma$-decays) such as Eq.~\labelcref{eq:17}, one cannot derive the related constraints to the QG coefficient $M_{D}/\zeta_{D}~(\simeq M_{\mathrm{QG}})$. Thus there is, essentially, \textit{no} incompatibility between these very strong $\gamma$-decay limits with our main result~\labelcref{eq:10}~(or \labelcref{eq:9}) just presented.

Following that same logic, we shall see that similar arguments can be applied to those studies on vacuum birefringence, which is normally absent but can be motivated by the reflexion symmetry~(parity/CPT) violation in Eq.~\labelcref{eq:5}. For photons with such \textit{birefringent} dispersion relations, the velocity variation of two polarization modes~($h_{\pm}$) is given by
\begin{equation}\label{eq:18}
\Delta v\left(\mathcal{E},h_{\pm}\right)=\mp\mathcal{E}/E_{\mathrm{LV}}^{(\mathrm{bire})},
\end{equation}
where we write explicitly the LV sign factor, and with also the requirement that the subluminal and superluminal parts of~\labelcref{eq:5} induce the same amounts of variation, as commonly assumed in some field-theoretic LV models. Here $E_{\mathrm{LV}}^{(\mathrm{bire})}$ governs the magnitude of CPT-odd LV effects, such as the helicity dependence of the group velocity. Strong constraints on $E_{\mathrm{LV}}^{(\mathrm{bire})}$ have been set in astroparticle physics by wide studies on polarimetric measurements~\cite{Shao:2011uc} from various sources over years. For instance, the observation of polarized lights from the Crab Nebula~\cite{Maccione:2008tq} has constrained the birefringence propagation in Eq.~\labelcref{eq:18} to the level $E_{\mathrm{LV}}^{(\mathrm{bire})}>1.3\times 10^{28}~\mathrm{GeV}$ at $2\sigma$, and the most stringent constraint comes from GRB~061122~\cite{Lin:2016xwj} with the relevant scale determined to be $E_{\mathrm{LV}}^{(\mathrm{bire})}\gtrsim 1.2\times 10^{35}~\mathrm{GeV}$, tightening by seven orders of magnitude the Crab Nebula constraint, and strikingly, by about thirteen-order over the earliest result~\cite{Gleiser:2001rm} from the radio galaxy 3C~256. These previous limits on the relative speed difference~\labelcref{eq:18} almost rule out LV-motivated birefringence, indicating that photon propagation in vacuum should \textit{not} be birefringent. Again we see that this coincides exactly with the last but not least prediction~\textbf{(iv)} of the space-time foam model. Specifically, since the coupling of the photon to the quantum D-brane foam is independent of photon polarization vector, as mentioned above, photons with opposite `helicities' but the same frequency would propagate at an identical velocity with $M_{\mathrm{QG}}$-suppressed variation from the absolute speed of light
\begin{equation}\label{eq:19}
\Delta c_{g}(\mathcal{E})\simeq -n_{D}g_{s}\mathcal{E}/M_{s},\ \text{for states}~~h_{+}, h_{-},
\end{equation}
and, as a result, the usual degeneracy of light speed amongst polarizations is retained while Lorentz invariance does not hold. This fact implies that those polarimetric observation cannot be used to bound the light velocity deformation~\labelcref{eq:3} emerged in the D-foam model. Thus those tight restrictions to $E_{\mathrm{LV}}^{(\mathrm{bire})}$ in Eq.~\labelcref{eq:18} cannot translate to upper limits for the quantity $n_{D}g_{s}/M_{s}$ entailed in the theory~(see Eq.~\labelcref{eq:19}), thereby, again, producing \textit{no} incompatibility with that given in Eqs.~\labelcref{eq:9} and~\labelcref{eq:10}.

Actually, from another point of view, since the LV photon decay and vacuum birefringence are absent in the D-particle foam model, it is very natural to expect that no observation would favor these two distinct phenomena to take place for photons, therefore, any attempts to search for these LV phenomena using high-energy $\gamma$-ray measurements would necessarily result in very tight constraints to certain types~(\textit{i.e.}, superluminal type and/or birefringent type) of Lorentz violation. Obviously, this is exactly the situation for current observational results, such as those strong astrophysical constraints, as discussed, including also our lower limit~\labelcref{eq:17} for photon decay, based on the latest near-PeV event detected by the Tibet AS array.

Hence, we conclude that the current \textit{non-observation} of superluminal photon decay phenomena and/or detectable birefringence effects can therefore serve as complementary supports for the theory discussed at the beginning.

It is also worth mentioning that, with the improvement of thirteen orders of magnitude over the past two decades, we could expect to have extra improvements on constraining possible vacuum birefringence. The situation for those effects due to superluminal Lorentz violation~(\textit{e.g.}, $\gamma$-decays) would be significantly improved as well in the future thanks to HAWC, the Tibet AS$\gamma$, and the new-generation cosmic-ray observatory LHAASO~\cite{Cao:2019lnn}, which should be able to perform more precise measurements of ultra-high-energy cosmic photons that could be used to strengthen our constraint~\labelcref{eq:17} at present from the AS$\gamma$ experiment. If more stringent bound can be cast indeed with the future observation, then this would be a strong sign of the absence of these LV phenomena, as indicated by the theory~(see predictions~\textbf{(iii)} and~\textbf{(iv)}), for reasons explained above, and therefore could serve as strong support for the theory.

We close by recalling that the linearly energy-dependent light velocity deformation which is subluminal but not birefringent emerges from a Type-IA string theory model for space-time foam. Given that high-energy astrophysical observations provide sensitive probes of such quantum-gravitational fluctuations in the velocity of light, we suggest that the light speed variation with $E_{\mathrm{LV}}^{(\mathrm{sub})}\gtrsim 10^{17}~\mathrm{GeV}$ from GRB photons~\cite{Shao:2009bv,Zhang:2014wpb,Xu:2016zxi,Xu:2016zsa,Xu:2018ien,Liu:2018qrg} can serve as a support for such theory, while being compatible with other astrophysical results, available today, based on photon decays and/or birefringent effects. Meanwhile, we extract a novel constraint to superluminal Lorentz-violation scale from the Tibet AS$\gamma$ near-PeV event to $\sim 8.39\times 10^{23}~\mathrm{GeV}$, which turns out to be the tightest LV photon decay limit reported so far, improved by about 1 to 3 orders of magnitude than those previous values in Refs.~\cite{Martinez-Huerta:2016azo,Albert:2019nnn}. Again this result serves as a complementary evidence to support the D-particle/string foam model. We thus reveal the possibility that the propagation anomalies of photons in cosmic space might have a realistic string-theory origin. Such observations can also provide hints for an interesting and long-lasting conjecture that space-time may not be the same as we usually perceive, but might exhibit `foamy' structures at tiny scales, thus could provide hopefully key insight into the nature of our Universe.

%
%

\section*{Acknowledgements}


We are grateful for useful discussions with Tianjun Li. This work is supported by National Natural Science Foundation of China (Grant No.~12075003).

\section*{References}

\end{document}